\documentclass[10pt,journal,compsoc]{IEEEtran}

\ifCLASSOPTIONcompsoc
  \usepackage[nocompress]{cite}
\else
  \usepackage{cite}
\fi

\usepackage[T1]{fontenc}
\usepackage{amsmath}

\usepackage[cmintegrals]{newtxmath}
\usepackage{amsfonts}

\usepackage{cite}
\usepackage{amssymb}
\usepackage{adjustbox}
\usepackage{multirow}
\usepackage{textcomp}
\usepackage{xcolor}
\usepackage{enumitem}
\usepackage{balance}

\PassOptionsToPackage{hyphens}{url}
\usepackage{hyperref}
\hypersetup{
    colorlinks = false,
    hidelinks = false
}
\usepackage{url}

\usepackage{algorithm2e} 
\newcommand{\nextnr}{\stepcounter{AlgoLine}\ShowLn}
\makeatletter
\newcommand{\removelatexerror}{\let\@latex@error\@gobble}
\makeatother

\usepackage{graphicx}
\usepackage{array}
\usepackage{textgreek}

\makeatletter
\newcommand\footnoteref[1]{\protected@xdef\@thefnmark{\ref{#1}}\@footnotemark}
\makeatother

\usepackage{tikz}
\usetikzlibrary{fit}
\newcommand*{\pie}[1]{
\trimbox{-2pt 0cm 0pt 0pt} {\begin{tikzpicture}[scale=0.85,baseline=-2]%
    \draw (0,0) circle (1ex);
    \fill[fill opacity=0.88,fill=black] (0,0) -- (90:1ex) arc (90:90-#1*3.6:1ex) -- cycle;
    \end{tikzpicture}
}} 
\newcommand*{\twopie}[1]{%
\trimbox{0pt 0cm 0pt 0pt} {\begin{tikzpicture}[scale=0.65,baseline=-2]%
    \draw (0,0) circle (1ex);%
    \fill[fill opacity=0.88,fill=black] (0,0) -- (90:1ex) arc (90:90-#1*3.6:1ex) -- cycle;%
    \end{tikzpicture}%
}} 

\usepackage{pifont}
\newcommand{\cmark}{\ding{51}}
\newcommand{\xmark}{\ding{55}}

\newcommand{\scozzertracer}{{\sc {SNIPER}}}

\newcommand\chal[1]{{\bf [C#1]}}

\begin{document}

\title{Designing Robust API Monitoring Solutions}

\author{
Daniele~Cono~D'Elia,
Simone~Nicchi,
Matteo~Mariani,
Matteo~Marini, and
Federico~Palmaro%
\IEEEcompsocitemizethanks{%
\IEEEcompsocthanksitem D.C. D'Elia, S. Nicchi, M. Mariani, and M. Marini are with Sapienza University of Rome, 00185 Roma, Italy (e-mail: delia@diag.uniroma1.it).
\IEEEcompsocthanksitem F. Palmaro is with Prisma, 00142 Roma, Italy.}%
\thanks{\bf This work has been submitted to the IEEE for possible publication. Copyright may be transferred without notice, after which this version may no longer be accessible.}%
\thanks{(Corresponding author: D.C. D'Elia.)}%
\thanks{Digital Object Identifier: none available at the moment}}

\IEEEtitleabstractindextext{%
\begin{abstract}
Tracing the sequence of library and system calls that a program makes is very helpful in the characterization of its interactions with the surrounding environment and ultimately of its semantics. Due to entanglements of real-world software stacks, accomplishing this task can be surprisingly challenging as we take accuracy, reliability, and transparency into the equation.
%
To manage these dimensions effectively, we identify six challenges that API monitoring solutions should overcome and outline actionable design points for them, reporting insights from our experience in building API tracers for software security research.
We detail two implementation variants, based on hardware-assisted virtualization (realizing the first general-purpose user-space tracer of this kind) and on dynamic binary translation, that achieve API monitoring robustly. We share our SNIPER system as open source.%
\end{abstract}

}



\maketitle

\IEEEpeerreviewmaketitle

\setcounter{footnote}{0}


\section{Introduction}
\label{se:introduction}


A modern operating system (OS) typically comes with large, heterogeneous software components that developers can build on when writing a software program. Operating systems expose their functionalities through Application Programming Interfaces (APIs) that compiled code accesses through well-defined prototypes and calling conventions.

The sequence of APIs that a piece of code may invoke during its execution characterizes its externally observable behavior and ultimately its semantics. Hence, monitoring API calls is a popular practice in dynamic analysis scenarios.

This operation is useful, for instance, in malware analysis and code reverse engineering activities in order to track how an untrusted piece of software interacts with the surrounding environment~\cite{Egele-CSUR08}. It is similarly valuable in dependability research for run-time monitoring~\cite{Estrada-EDCC15} and troubleshooting~\cite{patent} of programs, among others.

As the monitoring process requires an interposition mechanism for API calls, it also goes under the colloquial name of {\em API hooking}.
Researchers and practitioners may also resort to specialized forms of API monitoring, tailoring their implementations to different contexts. 

For instance, a malware sandbox monitors interactions with the operating system to sketch the high-level behavior of a sample under analysis. Sandboxes prevalently interpose on system calls~\cite{bluepill-tifs} so as to collect in a single spot events of interest originating from high-level APIs. This rationale is mainly justified by the desire not to miss behaviors exercised via unaccounted-for APIs, since to accomplish some task malware authors may sometimes draw from many alternative library APIs and use them in unexpected ways.

Conversely,  when moving to the manual dissection stage for notable samples~\cite{Cooperative-CYCON13}, malware analysts resort to monitoring solutions that trace also library APIs in order to understand \textit{how} a sample achieves some functionality. Consider a code fragment using an HTTP-related API: intercepting a packet transmission down in the software stack is not as informative as logging the API that originated it. Also, there are APIs whose uses cannot be directly inferred from the sole observation of one or more system calls.

Tracing high-level facts is in general valuable for many monitoring, troubleshooting, and reverse engineering tasks.
Notwithstanding the relevance and the many applications of API monitoring, we note that prior literature touches only slightly the design space and the {\em accuracy}, {\em reliability}, and {\em transparency} dimensions behind this task in the general case.

This is possibly reflected in how mainstream API monitoring systems available today fall short in one or more of these respects. Common flaws that we observed in our experience are incomplete argument tracing, missed calls to APIs that a program solves dynamically, and hooking artifacts that adversaries routinely identify. Also, most systems overwhelm users with calls that originate within the implementation of a high-level API: their logging does not bring users any actionable information, but only describes how the operating system realizes some functionality.


In the first part of this article we identify six key challenges towards accurate, reliable, and transparent API monitoring. We analyze the design space for a general-purpose API tracing system that can address them and work over different instrumentation technologies, with special attention to implications from security uses.

We then present two implementations of our design, detailing common traits and distinctive features of both. One builds on dynamic binary instrumentation and can operate standalone or as a support library for existing dynamic analysis systems based on this popular abstraction. The other builds on CPU virtualization extensions for more efficient and transparent instrumentation, and represents the first open solution of this kind. Following an extensive validation, we experimentally analyze their capabilities in tracing real-world programs and complex malware.

Our implementations target the Windows platform, covering a large collection of libraries and system calls, and can be extended to other systems. Although general-purpose, we incubated them as part of our malware analysis research: the tricky patterns found in this realm combined with quirks of Windows internals have been a tough training ground for their development.

We make the code of our API tracing system \scozzertracer{} available at \verb|https://github.com/dcdelia/sniper|.

%





\section{Background}
\label{se:background}
This section covers the main traits and underpinnings of the API handling process for Windows programs and libraries, and compares instrumentation technologies available to date for implementing an API monitoring system.

\subsection{Windows API Resolution and Internals}
\label{ss:winapis}

Windows programs can access functionalities of the surrounding environment by loading functions from \textit{dynamic-link library} (DLL) modules. In order to locate addresses for external function symbols, the Portable Executable format provides for an \verb|.idata| section for their listing. This section contains an array called \textit{import address table} (IAT) that the Windows loader populates at run-time with pointers to the desired functions, importing them from known DLLs.


DLL modules come with an analogous \verb|.edata| section for their public functions and storage, known as {\em exports}. Each entry of the {\em export address table} (EAT) is a relative virtual address (RVA) that is the offset of the export from the base address of the module. For an executable importing a function from a DLL, the loader will typically populate the involved IAT entry by looking up the RVA of the export in the EAT of the DLL and adding it to the base address that the system chose for the DLL when loading it.

However, there are alternative methods to locate API addresses. A program may manually load a DLL and retrieve its exports using the \verb|GetProcAddress| Windows API that does not touch the IAT. Furthermore, regardless of how DLL loading happens, a program may solve symbols in a covert manner by manually navigating the code modules in memory and parsing their \verb|.edata|. This happens frequently with software wrapped by executable packers and protectors---popularly used by both malicious and benevolent programs~\cite{Egele-CSUR08}---and with shellcode.

When it comes to the internal structure of a DLL, an exported API can be of different kinds. The common case is when its logic is fully contained in the code starting at the given RVA. In other cases the code is partial and ends with a tail jump to another function (either private or exported) or to an export from another DLL. The latter type is frequent, for instance, in {\tt kernel32.dll} relying on {\tt kernelbase.dll}. In other cases the RVA does not point to code, but represents a {\em forwarder} export~\cite{forwarders}: this instructs the loader to silently rewire any IAT entry referencing it to point to another export from another DLL. Due to these factors, determining for an export what are the ``exit points'' to the caller function is not always immediate.


For API prototype information, the Windows SDK provides header files specifying for each API the calling convention~\cite{callingconv} (typically \textit{stdcall}) and the {\em input modifier} of each function argument, that is, when an argument identifies an input (IN) or an output (OUT) value, or both (INOUT)~\cite{inout}. An output argument is usually a pointer to a region where the API can write output data. Alongside prototype information, headers introduce a large number of type aliases and structure declarations.

System calls, or {\em syscalls} for short, operate in a different way. Classic programs access them through user-space wrappers available as exports of {\tt ntdll.dll} that set the syscall ordinal in register EAX and trigger a software interrupt. Adversaries can elude monitoring strategies that check such wrappers by extracting from {\tt ntdll.dll} the ordinals for the current Windows version and triggering  the interrupt with own code, realizing a so-called {\em direct syscall}.

Experienced writers may also use undocumented syscalls to complicate the analysis: prototypes for them may only be found in reverse engineering forums. 

\subsection{Instrumentation Technologies}
\label{ss:instrumentations}
As we will see throughout this article, the type of instrumentation a system uses to interpose on API calls affects several dimensions of the overall monitoring efficacy.

A possible avenue is to patch an IAT entry so as to point to a tracing stub that logs the call and in turn invokes the intended API. This approach is known as {\em IAT hooking} and unfortunately misses calls to APIs made without the IAT (\textsection\ref{ss:winapis}). A better alternative that covers also those is to place instrumentation inside API code, modifying the prologue of monitored functions with the insertion of a {\em trampoline} to a tracing stub. Mainstream commercial tracers follow either approach; a major weakness of both is that an adversary can clearly recognize the modifications they introduce.

Instrumentation artifacts are a well-known problem for dynamic analyses that operate through binary patching: for this reason, other technologies have gained popularity in security research~\cite{SoK-DBI}.

{\em Dynamic binary translation} (DBT) systems can trap execution at arbitrary instructions based on their type (e.g., control transfer instructions) or address, and provide the running code with the illusion that instrumentation is not present. A popular DBT technique for user-space monitoring of programs is {\em dynamic binary instrumentation} (DBI)~\cite{DynamoRIO-CGO03,Pin-PLDI05,SoK-DBI}. When an analysis has to deal with kernel-level or system-wide execution flows, researchers have used instead whole-system emulators like QEMU~\cite{QEMU-ATEC05} as DBT systems to instrument an entire virtual machine (VM). In such out-of-VM scenarios, {\em virtual machine introspection} (VMI) tools come into play to overcome the semantic gap~\cite{Virtuoso-SP11} arising when having to retrieve high-level features of the underlying OS and processes by accessing the raw memory of the VM.

The advent of CPU virtualization extensions like Intel VT favored new instrumentation primitives with better performance and transparency. By maintaining a split view of code and data pages in the Extended Page Table, the authors of SPIDER~\cite{SPIDER-ACSAC13} propose {\em invisible breakpoints} for registering analysis callbacks at physical page addresses. This design defeats introspective attacks: to insert a breakpoint, SPIDER creates a code page for instruction fetching and modifies it, leaving the original code untouched in a data page visible to read/write operations. The DRAKVUF~\cite{DRAKVUF-ACSAC14} sandbox uses a variant of this mechanism to hook kernel code for syscalls and few selected user-space APIs. However, as we will explain later in the article, lazy loading mechanisms and other OS entanglements get in the way when one wishes to use VT-based instrumentation to trace arbitrary user-space APIs from heterogeneous libraries.


\section{Design Space of API Monitoring Systems}
\label{se:design}

In this section we identify and analyze common problems behind API monitoring systems, discussing design alternatives that are not bound to specific instrumentation technologies. The possibilities we outline are readily actionable, as \textsection\ref{se:implementation} will detail for our DBI and VT-based implementations.


\subsection{Challenges in API Monitoring}
\label{ss:challenges}


We identified six challenges along the path to accurate, reliable, and transparent API monitoring solutions:

\begin{description}
\item[{[C1]} Transparency.] Adding probes or other instrumentation for intercepting calls to an API may introduce artifacts that an adversary can look for~\cite{Egele-CSUR08}, or collide with recent OS exploit mitigations against API hijacking~\cite{msdn-iaf}.
\item[{[C2]} Recall.] The points in the software stack where instrumentation takes place also determine how many of the actual calls a tracer can capture (as we have seen for instance with the limitations of IAT hooking in \textsection\ref{ss:instrumentations}).
\item[{[C3]} Coverage.] Tracing argument values for an API call is more informative than tracing API symbols alone. With an ample universe of library, a programmatic approach to extract prototypes and data type declarations is necessary to avoid incomplete information retrieval.
\item[{[C4]} Output values.] A tracer should be able to capture the return value of an API call, as well as any data the API wrote to output  locations supplied by the caller. 
\item[{[C5]} Relevant calls.] An API call in a program may lead to many intra- and inter-component API calls along the software stack: these {\em internal calls} bloat monitoring logs but are hardly informative and should be filtered out.
\item[{[C6]} Derived flows.] A tracer should cover derived execution flows like child processes and remote (i.e., created in other processes) threads. Adversaries may also use them to hide API calls from the analysis~\cite{APIChaser13}.
\end{description}

We observe that \chal{1} and \chal{6} are compelling aspects in many security and dependability settings, \chal{2-4} impact the soundness and completeness of high-level analyses that build on API monitoring, while \chal{5} affects the usability of a monitoring system when a human agent is involved.

We reviewed publicly documented API monitoring solutions and identified\footnotemark{} three products that are popular among security professionals and two DBT-based research systems. After careful analysis and testing, we found all of them to fall short in one or more of the six \chal{1-6} respects (Table~\ref{tab:tools}). We defer the discussion of each system to \textsection\ref{ss:comparison} so as to allow for a detailed comparison with our solutions. 

To come up with robust solutions for API monitoring, in this work we reason on the key design choices that impact these respects. The next sections will discuss possible alternatives (where applicable) and motivate our choices. We make an effort to pursue accuracy and reliability of the tracing without tying our design to an instrumentation technology and its transparency properties.

\subsubsection{Threat Model}
\label{sss:threat-model}
As we devote special attention to software security uses, we assume that the program under analysis may run introspective sequences to reveal API monitoring mechanisms typically found in malware and in programs shielded by executable protectors~\cite{csur19evasions}. These sequences may verify the integrity of its code and data (Test T1) or of DLL code in memory by using precomputed signatures or comparing it with its counterpart on disk (Test T2).  We craft an adversarial program for testing purposes \chal{1} that for T1 compares each IAT entry of the running program with the expected address for its symbol (found via the EAT of the DLL exporting it), and for T2 reads the DLL contents from disk to a buffer, applies relocations (to match where Windows loaded the real DLL), and compares the first 8 bytes of every imported DLL symbol with our copy. T1 will expose stubs for IAT hooking, while T2 will reveal hooking trampolines.

 
\begin{table}[t!]
\begin{center}
\begin{footnotesize}
\adjustbox{max width=\columnwidth}{
\begin{tabular}{|l|l|c|c|c|c|c|c|c|}
\cline{3-4}
\multicolumn{2}{c|}{} & \multicolumn{2}{c|}{\textbf{C1}} & \multicolumn{3}{c}{}  \\
\hline
\textbf{Tracing system} & \textbf{Technology} & \textbf{Test T1} & \textbf{Test T2} & \textbf{C2} & \textbf{C3} & \textbf{C4} & \textbf{C5} & \textbf{C6}\\
\hline
API Monitor v2{\textalpha}-r13 & IAT hooking & \xmark & \cmark & \pie{40} & \pie{70} & \pie{100} & \pie{0} & \pie{0} \\
SpyStudio v2.9.2 & Trampolines &  \cmark & \xmark & \pie{40} & \pie{40} & \pie{100} & \pie{0} & \pie{0} \\
WinAPIOverride 6.6.6 & Trampolines &  \cmark & \xmark & \pie{70} & \pie{70} & \pie{100} & \pie{0} & \pie{40} \\
\hline
drltrace~\cite{drltrace} & DBI &  \cmark & \cmark & \pie{100} & \pie{40} & \pie{0} & \pie{70} & \pie{40} \\
PyREBox~\cite{MarianoBox} & QEMU-TCG & \cmark & \cmark & \pie{100} & \pie{100} & \pie{100} & \pie{0} & \pie{70} \\
\hline
\scozzertracer{} & DBI, VT & \cmark & \cmark & \pie{100} & \pie{100} & \pie{100} & \pie{100}  & \pie{100} \\
\hline
\end{tabular}
}
\end{footnotesize}
\vspace{3pt}
\caption{Mainstream tools and research systems with generic API tracing capabilities. Circles are filled to indicate if an aspect is met to a basic, good, or optimal extent (details discussed in \textsection\ref{ss:comparison}).\label{tab:tools}}
\end{center}
\end{table}

\footnotetext[1]{We involved in the selection an independent malware analyst with ten years' experience as principal malware analyst in security firms.}
\footnotetext[2]{And hand-written headers for uncounted structures and syscalls.}

\subsection{Scope of Monitoring} 
\label{ss:design-scope}

Two elements help to determine the scope of an API tracer: the breadth and level of detail of available prototype information for  APIs, and the treatment of internal calls.

\subsubsection{Prototypes}
Windows operating systems offer dozens of DLLs to programmers. Each DLL file comes only with information for symbols and relative locations of its exported functions (\textsection\ref{ss:winapis}). A sound way to obtain argument information \chal{3-4} is to cross-reference DLL symbols with function declarations from the header files of the Windows SDK\footnotemark: the Deviare engine~\cite{deviare2} offers an infrastructure to this end.

This approach is general and can be applied also to third-party libraries if their headers are available. A programmatic extraction shall include the size of each argument so to be able to fetch their values at run-time; pointer types require a recursive valuation. However, input modifiers are not present in headers and can be retrieved from the MSDN documentation using a crawler like~\cite{zynamics}. 


\subsubsection{Relevant Calls}
We reasoned on what makes a traced call informative for a user \chal{5}. Internal calls happening within one or more DLLs do not provide valuable insights to the user, but only describe how the OS implements the outer high-level API that triggers them. Eventually, they make up a large fraction of tracing logs.
%
To spot them, we define the scope of a call to be {\em relevant} only when the call is made in code belonging to the program under analysis and eventually returns to it.

This definition rules out internal calls or jumps to other exports, and redirections within API code (\textsection\ref{ss:winapis}), and captures syscalls that programs invoke directly from their code. To cope with \chal{6}, by {\em program under analysis} we mean the process where the code first runs, the child processes and remote threads it creates, and any recursive byproduct.

\subsection{Hook Insertion}
\label{ss:design-precision}

As part of its working, an API monitoring system must interpose on specific events: the invocation of an API ({\em API entry}) and the moment it returns to its caller ({\em API exit}). The placement of hooks through instrumentation impacts the recall of a tracer \chal{2} and the call argument extraction \chal{3,4}.






\subsubsection{API Entry Events} 
\label{sss:api-entry-events}
Static code analyses face soundness and completeness problems for locating where an arbitrary program may invoke an API; obfuscation and anti-analysis measures only make this problem harder~\cite{Egele-CSUR08}. IAT hooking (\textsection\ref{ss:instrumentations}) is similarly unfit for hooking purposes as it struggles with recall \chal{2}.
A reliable choice to intercept API entry events is instead to monitor dynamically when execution reaches code from an API.

One possibility is to interpose on control transfer instructions. For instance, PyREBox hooks every {\verb|call|} (optionally {\verb|jmp|} and {\verb|ret|} too) instruction in the execution and checks its target against a list of addresses for monitored APIs. Unfortunately, this approach introduces unnecessary overhead for control transfers unrelated to APIs, which are dominant in practice. It also requires an instrumentation facility that can hook instructions by type as the CPU sees them during fetching, ruling out static rewriting (self-modifying code breaks it~\cite{SoK-DBI}) and VT-based instrumentation.

We argue instead for placing instrumentation in the prologue of an API's code: interposition will thus happen only and every time the API function is invoked, either by the program or internally down the software stack.

To locate suitable addresses, we target every unique RVA that appears in the EAT of a loaded DLL and is not a forwarder. For forwarders we will instrument the API hosting its actual code. This design maximizes recall \chal{2}.

\subsubsection{API Exit Events}
\label{sss:api-exit-events}
\begin{figure}[t!]
\begin{center}
\includegraphics[width=0.98\linewidth]{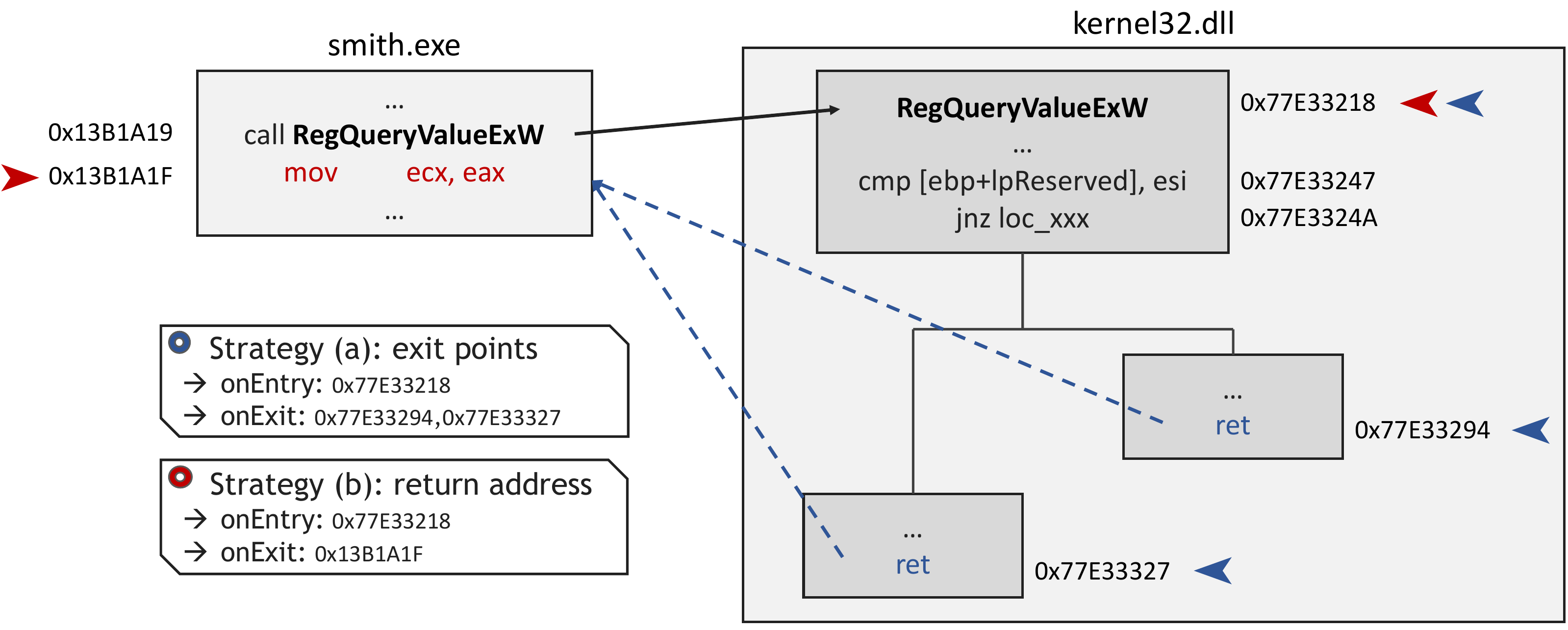}
\end{center}
\vspace{-3.5mm}
\caption{API call handling with strategy (a) or (b) for exit events. Arrows placed next to instruction addresses represent hooks.\label{fig:retaddr-exitpoints}}
\end{figure}

Intercepting when execution returns to the caller of an API is necessary to log return values and output arguments \chal{4}. Figure~\ref{fig:retaddr-exitpoints} depicts two viable options to this end:
\begin{enumerate}[label=(\alph*)] 
\item at DLL load time, placing hooks on the one or more exit points for the code that implements the API;
\item at API entry time, placing a hook on the instruction that represents the return address for the call.
\end{enumerate}


Strategy (a) of chasing exit instructions is not immediate to pursue due to the redirections present in many API implementations (we discussed them in \textsection\ref{ss:winapis}). Based on the insights from analyzing Windows DLLs, we wrote a static analysis that processes partial implementation stubs and tail calls in APIs to determine their exit points. We found exotic cases where an export makes a tail call to an export from another DLL that in turn leads to a tail call to another export from a third DLL.

Strategy (b) of chasing return addresses looks easier at first, but the hooking logic should be carefully designed to process only a real API exit. In fact, the instruction following the call may be a join point in the control flow graph later reached from basic blocks that do not end with an API call. We found many instances of this pattern for example in Microsoft utilities (details in \textsection\ref{apx:a}).

Unless a program contains a pathological number of these patterns, strategy (b) is more attractive as it brings fewer invocations of the analysis callback for API exit events, as our design discards internal calls. In fact, strategy (a) would trigger the callback also for them and analysis code must ignore them. Choosing one scheme over the other has no impact on the other components of the design, but depends mainly on the capabilities and efficiency of the chosen instrumentation technology: we detail this aspect in \textsection\ref{ss:impl-exit}, explaining why and when the two may also profitably coexist when implementing an API tracer. 


\subsubsection{Argument Extraction}
To support the retrieval of input and output arguments for an API call, the instrumentation facility should expose CPU register values and program memory to the tracer.

Upon API entry events, the stack pointer value suffices to locate the arguments under 32-bit {\em stdcall} (Windows) and {\em cdecl} (UNIX-like systems) calling conventions. 64-bit code requires accessing dedicated registers for the first 4 arguments, then any additional one is passed on the stack. Similar considerations apply to floating-point arguments, which we omit in the remainder of the article for brevity.

For exit events, the return value is available in register EAX, plus EDX for wider data types. 64-bit output arguments passed in registers can be saved safely during the entry event. Prototype information is essential on API entry and exit events for computing offsets for stack arguments by taking into account their size and order.


\section{Our Tracing Solutions}
\label{se:implementation}
\begin{figure}[t!]
\begin{center}
\includegraphics[width=0.99\linewidth]{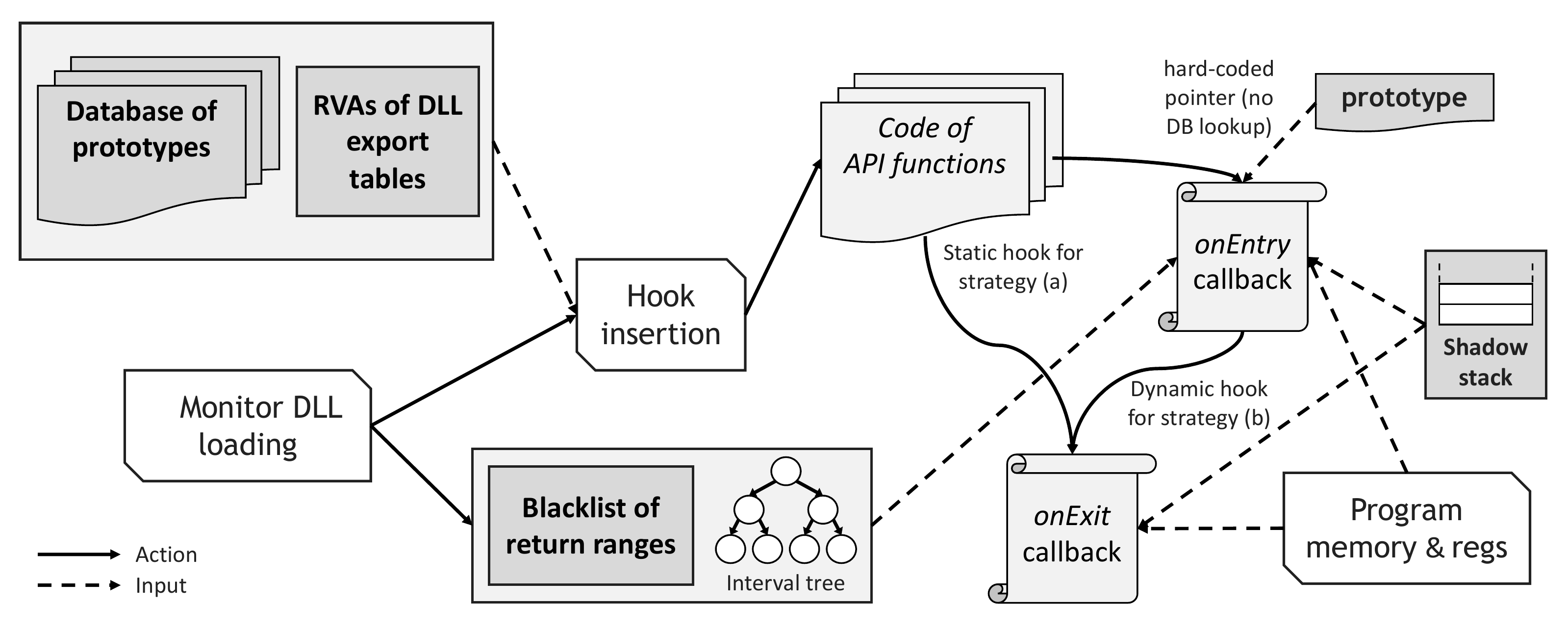}
\end{center}
\vspace{-3.5mm}
\caption{Bird's eye-view of the proposed \scozzertracer{} system.\label{fig:design}}
\end{figure}

Figure~\ref{fig:design} portrays the architecture of our \scozzertracer{} system, which embodies the design points emerged from the discussions of \textsection\ref{se:design}.
We detail next its DBI and VT-based implementations in their common traits and distinctive features.

\subsection{Instrumentation Technologies} 
The design options outlined throughout \textsection\ref{se:design} are general, in the sense that they can be implemented using different instrumentation technologies. However, choosing one technology over others can lead to transparency concerns \chal{1}.

We implement the first variant of \scozzertracer{} in Pin~\cite{Pin-PLDI05}, a popular DBI choice in programming language, software testing, and security research. The variant ships as a high-level library suitable either for standalone usage as an in-guest API monitoring tool or for being plugged in existing analysis systems based on Pin, which are numerous in security research~\cite{SoK-DBI} as well as in other communities.

The DBI abstraction ensures that every address or value manipulated by the program matches the one expected in a native execution. Under the hood, Pin operates by JIT-compiling and instrumenting code in a designated cache area: any hook we insert will not be visible to introspective attempts from the threat model of \chal{1} (\textsection\ref{sss:threat-model}). We also augment Pin with recent mitigations for DBI artifacts~\cite{SoK-DBI}. These factors contribute to making our tracer less conspicuous than commodity systems based on binary patching (\textsection\ref{ss:instrumentations}). 



The second variant brings a new piece to the research landscape: a tracer compatible with modern designs for efficient out-of-VM analyses via VT extensions. This variant is particularly suitable for scenarios where minimal invasiveness is desirable (e.g., with code sensitive to environment artifacts~\cite{DRAKVUF-ACSAC14} or slowdowns~\cite{Bruening-VEE12}) and for monitoring system-wide flows. For placing hooks we use the implementation of invisible breakpoints (\textsection\ref{ss:instrumentations}) from DRAKVUF~\cite{DRAKVUF-ACSAC14}, a whole-system analysis framework based on the Xen hypervisor. 

\subsection{Relevant Calls and Execution Units}
\label{ss:relevant-calls-units}
In \textsection\ref{ss:design-scope} we mentioned the advantages of restricting tracing to calls explicitly made by the program under analysis \chal{5}. The first dimension of the problem is to identify processes and threads relevant to this end \chal{6}, which requires different provisions for different instrumentation technologies.

Pin naturally operates on a single process of interest, and offers APIs to intercept the creation of a child process or a remote thread in an existing process~\cite{SoK-DBI}: we use them to extend the instrumentation to such flows automatically.

\smallskip
In the VT-based scenario the subject of instrumentation is the entire system, hence an invisible breakpoint in API code may be triggered by processes unrelated to the one(s) of interest if they try to use that API. We carry out a bookkeeping work to identify relevant execution units: we wrote a VMI component that, starting from a process under analysis, tracks the creation of child processes and remote threads recursively. The component maintains a pool of IDs for monitored threads and their enclosing processes.

To this end, we hook the kernel code of syscalls \verb|NtCreateThreadEx| and \verb|NtTerminateThread| when they return to user mode. For thread creation we read the value of the output \verb|ThreadHandle| argument of the syscall: since it is valid only for the caller, we look into the Windows Object Manager to locate the referred \verb|_ETHREAD| data structure and extract from there the process ID (different than the caller's when dealing with remote threads) and the thread ID. For thread termination we discriminate whether the caller thread is terminating itself or another thread, accessing the related \verb|_ETHREAD| data in the latter case.

With those IDs we maintain and update the above-mentioned pool: when execution hits an invisible breakpoint from an API hook, the raised callback checks whether the current execution unit belongs to the pool. In case of code injection patterns, this design also lets us ignore activities from the ``authentic'' threads of a victim process.

\medskip
To identify internal calls happening in Windows components \chal{5} we can then check, under any instrumentation, in which region the return address of an API call falls.

A whitelisting approach that logs only calls that return to code regions belonging to the program can be a slippery road: not only malware and protected executables, but even COTS programs can exercise exotic behaviors like executing code from the heap or change section permissions~\cite{Bruening-VEE12}.

We found it safer to build a {\em blacklist of return ranges} for calls to discard, and populate it with code section ranges of Windows DLLs (such addresses would indicate an internal call) as those get loaded and unloaded. This scheme turns out to be robust and efficient: as intervals are disjoint, we use an interval tree with logarithmic lookup cost. We complement the range lookup operation with ad-hoc measures for DLL tail jumps (\textsection\ref{ss:winapis}) that we present in the next section. 


\subsection{Hook Insertion and Callbacks}
\label{ss:hooks}

In the following we detail the registration and functioning of analysis callbacks for logging API calls from program code along with their input arguments and output values.

\subsubsection{API Entry}
\label{sss:hook-api-entry}
As motivated in \textsection\ref{sss:api-entry-events}, we target the RVAs of the symbols exported from loaded DLLs in order to place hooks for achieving interposition on API entry events. 

\footnotetext[3]{The authors of PyREBox released a large one that we use and refine in a few aspects, e.g., to distinguish INOUT arguments from OUT ones.}

DBI engines offer facilities to intercept loader activities. Once we identify the base address of a DLL module of interest, we locate its EAT, cross-reference the names of exported functions with a database of prototypes\footnotemark, and compute the run-time addresses of functions using their RVAs. We instrument the first instruction in each function and register an {\em onEntry} analysis callback, hard-coding the address of the prototype information for the API as argument for it so to avoid an expensive run-time lookup. This approach is independent of the Windows version in use, and has performance advantages as we can use Pin's IMAGE mode to place efficient ahead-of-time instrumentation~\cite{pin-manual}.

\smallskip
In the VT-based scenario we can equally parse EATs for RVAs or load precomputed Rekall profiles~\cite{rekall} for the current Windows version. Similarly to the DBI case, we insert an invisible breakpoint at the first instruction in each function, associating to it an {\em onEntry} analysis callback with the address of its prototype information as hard-coded argument. The difference with the other variant is that the callback will first determine whether the intercepted thread belongs to the pool of threads we wish to monitor.

Following the approach of SPIDER~\cite{SPIDER-ACSAC13}, an invisible breakpoint takes the form of an \verb|int 3| sequence replacing the original instruction: when hit, the hypervisor intercepts the raised \#BP exception, invokes the callback, executes the original instruction in single-step, and re-enables the breakpoint. As we anticipated in \textsection\ref{ss:instrumentations}, de-synchronization of data and code views for pages hides the presence of such a breakpoint from an adversary.

Invisible breakpoints operate on physical pages: adding instrumentation to logical addresses requires their translation to physical ones. Unfortunately, as Windows implements a lazy loading mechanism, when we intercept a DLL loading event not all its pages may be amenable to hook insertion: put in other words, for a given logical address there might be no physical page yet~\cite{drakvuf-issues}.

We wrote a component that loads DLLs of interest in a separate process and reads code from their sections forcing page materialization: since such pages are normally shared among processes, we can place hooks also for the program of interest. This scheme still misses a few corner cases, but luckily other researchers concurrently developed a mechanism to force page faults and materialize pages upon DLL loading~\cite{drakvuf-icedevml}, using a new feature of libvmi~\cite{libvmi} that meanwhile became available. We have started to extend our implementation to integrate their technique.

\medskip
\paragraph*{\em onEntry Callback} This analysis routine takes as input the ESP register value (to access the return address and stack arguments), a pointer to the prototype information for the current API, and further register values where needed (e.g., with 64-bit code). We provide a simplified pseudocode in Figure~\ref{alg:stack} that is agnostic to the instrumentation technology. 

We maintain a (thread-local) {\em shadow stack} of currently monitored functions\footnotemark{}. Line~\ref{algline:restrict} restricts logging to calls made in user code, or we would be mirroring also API calls happening within DLLs. Line~\ref{algline:internal} discards internal jumps and tail calls to other exported functions, which would see the same return address in program code of their caller (current top stack entry). Line~\ref{algline:ra} deals with hooking the return address when we use strategy (b) for handling API exit events. Line~\ref{algline:stale} performs sanitization of stale stack frames in case of instrumentation glitches: if ESP is at higher addresses than the ones stored, those calls likely returned already since the stack grows downwards to lower addresses. Finally, lines~\ref{algline:entry-push} and~\ref{algline:entry-parse} update the shadow stack and log the call, respectively.

\begin{figure}[t!]
\begin{footnotesize}
\removelatexerror
\begin{algorithm}[H]
  \DontPrintSemicolon
  \SetAlgoNoEnd
  \SetAlgoNoLine
  \SetNlSkip{-0.45em}
  \SetKwFunction{FMain}{onEntry}
  \SetKwProg{Fn}{function}{:}{}
  \Fn{\FMain{threadID, ESP, prototype, ...}}{
  \Indmm
  	\nextnr \nllabel{algline:restrict}\lIf{*ESP $\in$ RangeBlacklist}{return}
	\nextnr SStack = getTLS(threadID) //$\,\textit{thread-local storage}$\;
	\nextnr \If{\textsf{\bf not} SStack.empty()}{\Indmm
		\nextnr cInfo = SStack.peek() //$\,\textit{recorded call information}$\;
		\nextnr \nllabel{algline:internal}\lIf{*ESP == cInfo.ra \&\& ESP == cInfo.esp}{return}}
  	\nextnr \nllabel{algline:ra}hookReturnAddress(*ESP) //$\,\textit{skip under strategy (a)}$\;
	\nextnr \nllabel{algline:stale}removeStaleEntries(SStack, ESP) //$\,\textit{from top while cInfo.esp $\le$ ESP}$\;
   	\nextnr \nllabel{algline:entry-push}SStack.push($<$*ESP, ESP, prototype$>$) //$\,\textit{$<$ra, esp, proto$>$ call info}$\;
	\nextnr \nllabel{algline:entry-parse}parseArgsOnEntry(ESP, prototype, ...)\;      
  }
  \vspace{3.5mm}
  \SetKwFunction{FMain}{onExit}
  \SetKwProg{Fn}{function}{:}{}
  \Fn{\FMain{threadID, ESP, EIP, EAX, ...}}{
  \Indmm
	\nextnr SStack = getTLS(threadID)\;
	\nextnr \nllabel{algline:emptystack}\lIf{SStack.empty()}{return}
	\nextnr idx = SStack.size() - 1\;
	\nextnr cInfo = SStack[idx]\;
	\nextnr \While{true}{\Indmm
	\nextnr \nllabel{algline:exit-ra-cmp}\lIf{cInfo.ra == EIP $\parallel$ -\,-idx $<$ 0}{break}
	\nextnr cInfo = SStack[idx]}
  	\nextnr \nllabel{algline:negativeidx}\lIf{idx == -1}{return}
	\nextnr \nllabel{algline:exit-esp-cmp}\If{ESP == cInfo.esp + 4 + cInfo.prototype.retN}{\Indmm
		\nextnr \nllabel{algline:exit-parse}parseArgsOnExit(cInfo.esp, cInfo.prototype, EAX, ...)\;
		\nextnr SStack.resize(idx) //$\,\textit{pops one or more elements}$
	}
  }
\end{algorithm}
\end{footnotesize}
\vspace{-1.5mm}
\caption{Analysis callbacks executed upon API entry and exit events.\label{alg:stack}}
\vspace{-2mm}
\end{figure}

\subsubsection{API Exit Events}
\label{ss:impl-exit}
In \textsection\ref{sss:api-exit-events} we presented two strategies for tracing API returns: hooking exit instructions (a) or return addresses (b). For strategy (a) we place hooks at DLL load time at the exit points identified with static analysis, while for (b) we place them dynamically upon API entry events (line~\ref{algline:ra}).



We implement strategy (a) in Pin using the IMAGE mode as with entry events. For strategy (b) Pin lacks a neat way to place callbacks on instruction addresses during execution without resorting to heavy-duty features like TRACE mode (and its ROUTINE mode is unreliable for exit events~\cite{pin-manual}). We build instead on the recently introduced {\em PIN\_RemoveInstrumentationInRange} primitive to force Pin to recompile and reanalyze only the instruction at the return address. Recompilation is required only the first time a new return address is hit at line~\ref{algline:ra}. As we will explain shortly, the {\em onExit} callback ignores subsequent spurious raises of the hook, so we are not forced to unregister the hooking probe once the handling of an exit event completes.

In the VT-based scenario invisible breakpoints naturally back both strategies, as they can target arbitrary addresses.



\footnotetext[4]{We say {\em functions} as in a thread the concurrently active functions that return to user code may be multiple: this happens for instance when a program loads a custom DLL (hence we equate it with program code that we seek to monitor) with {\em LoadLibrary} and its {\em DllEntryPoint} function invokes one or more APIs before {\em LoadLibrary} returns.} 

\medskip
In \textsection\ref{sss:api-exit-events} we mentioned that strategy (b) reduces the fraction of times {\em onExit} is invoked for uninteresting events that have to be discarded. However, there could be cases where it may be more intrusive for the program under analysis (e.g., due to recompilation events in Pin), or an adversary knowing the details of the system may tamper with return addresses on the stack. \scozzertracer{} retains support for both schemes, which can coexist seamlessly and profitably.

In more detail, let us consider a new DLL or function for which we did not precompute exit points: we can instrument the return addresses for them, and use the other scheme for the rest of the APIs. Similarly, this flexibility helped us when developing the VT-based version, as DRAKVUF failed the breakpoint insertion at some exit instructions with no apparent reason, but we could fall back to hooking the return addresses for those APIs.


\paragraph*{\em onExit Callback}
%

The pseudocode in Figure~\ref{alg:stack} for the analysis routine for exit events is instrumentation-agnostic and embodies strategy (b). Initially, the routine looks up the most recent shadow stack entry that matches the current return address, which the instruction pointer EIP has just reached when the callback triggers.
%

It then invokes a routine for processing the return value and output arguments, passing to it the ESP value seen on entry and the prototype (both were saved to the shadow stack at that time) for the current API call, register EAX and where needed EDX for the return value, and any saved output arguments passed via registers for 64-bit code.

Lines~\ref{algline:emptystack} or~\ref{algline:negativeidx} can be hit when a previously hooked return address is reached by blocks that did not make an API call (\textsection\ref{sss:api-exit-events}, \textsection\ref{apx:a}). This logic is semantically equivalent to disabling instrumentation at the return address, which may not always be cheap (e.g., extra recompilations in Pin). A sanity check at line~\ref{algline:exit-esp-cmp} compares the current ESP value against the one stored by {\em onEntry} for the frame, ``undoing'' the effects of the \verb|ret N| instruction\footnotemark{}. We observed that in practice simply checking for $\textit{ESP}>\textit{cinfo.esp}$ is a reliable approximation of the condition. 


For strategy (a), since the callback would trigger on an exit point, we would see that the instruction pointer EIP has not been diverted yet to the return address (which can be found however at *ESP), and that the stack pointer has not been adjusted with the displacement associated with the return sequence. We adapt the code for {\em onExit} from Figure~\ref{alg:stack} to this strategy by simply replacing EIP with *ESP at line~\ref{algline:exit-ra-cmp} and using $\textit{ESP} == \textit{cinfo.esp}$ as condition at line~\ref{algline:exit-esp-cmp}.

\subsubsection{Argument Extraction}
\label{sss:parameters}
The subroutines that {\em onEntry} and {\em onExit} call at lines~\ref{algline:entry-parse} and~\ref{algline:exit-parse}, respectively, are conceptually similar.

Both may have to locate data from the stack, computing offsets based on the size of each previous argument in the prototype. While logging primitive types is a straightforward task, with pointers we need to distinguish the type and size of a pointed object, but first of all verify whether a pointer is meaningful, that is, if it points to valid data.

Ideally, a sound way would be to take into account the API semantics (e.g., checking its return code to discriminate errors), but this may be unrealistic for a general-purpose tracer. We can check instead if the pointed object falls into valid memory and call a print helper for its contents.

This operation is immediate for fixed-size objects such as primitive types or structures. For variable-size objects like strings we cannot rely on the presence of some terminator when an API fails: we conservatively fetch a predefined amount of bytes from the address, reducing it if the chunk would span two pages with the second being invalid.

\subsection{System Calls}
\label{ss:syscalls}
In the presentation we postponed the discussion of how our implementations address syscalls as there are some unique aspects to their handling.

From the program's perspective, syscalls are self-contained: they happen over a privilege mode change (to kernel mode when invoked, back to user mode upon termination). Hence, no shadow stack update is needed.

A good source of prototype information is the database from the DrSyscall module of DynamoRIO~\cite{DynamoRIO-CGO03}. We build on it as it covers many undocumented syscalls and provides auxiliary data needed to determine an argument type for cases where it depends on another argument of the syscall.

For syscall instrumentation, DBI and more generally DBT systems naturally intercept privileged sequences involved in their invocation and return~\cite{Pin-PLDI05}. In Pin we register two callback routines for syscall entry and exit events: from there we extract the syscall ordinal, use it as array index to retrieve the corresponding prototype from the DrSyscall database, and extract the current syscall arguments.

In the VT-based scenario we cannot interpose on instructions by type, so we follow the design proposed by the authors of DRAKVUF in~\cite{DRAKVUF-ACSAC14}, that is, we instrument the entry and exit instructions of the syscall implementations found in the NT kernel of Windows.

Finally, we deem a syscall relevant if it returns either to program code directly (as with direct syscalls found in malware) or to some {\em Nt} library wrapper from {\tt ntdll.dll} called in turn from program code. For the latter case we walk back the stack mimicking the effects of the epilogue instructions of the wrapper: we check if the stack frame of the method returns to program code or to an address in the DLL range blacklist, discarding the call in the second case.


\footnotetext[5]{After the instruction, the stack pointer value will be higher than the ESP value seen by {\em onEntry} by r+N bytes: r=4 for 32-bit code and 8 for 64-bit code, while N is the space used on stack in argument passing for {\em stdcall} APIs (we remind our readers that in {\em stdcall} functions the callee has to clean the stack for the caller) and zero for {\em cdecl} ones.} 



\newcommand*{\cpar}[2]{\parbox{#1}{\centering #2}}
\makeatletter 
\newcommand{\thickhline}{%
    \noalign {\ifnum 0=`}\fi \hrule height 1pt
    \futurelet \reserved@a \@xhline
}
\newcolumntype{"}{@{\hskip\tabcolsep\vrule width 1pt\hskip\tabcolsep}}
\makeatother

\begin{table*}[ht!]
\begin{adjustbox}{max width=0.99\textwidth}
\begin{footnotesize}
\setlength\extrarowheight{1pt}
\begin{tabular}{|c|r|r|r|r|r|c|c|c|c|c|c|c|c|c|}
\cline{2-14}
\multicolumn{1}{c|}{} & \multicolumn{2}{c}{\multirow{1}{*}{\bf \# of syscalls}} & \multicolumn{3}{|c|}{\bf \# of DLL calls} &  \multicolumn{3}{c|}{\bf DLL APIs (from program)} & \multicolumn{5}{c|}{\bf Avg call processing time (\textmu{}s)}  \\
\cline{2-14}
\multicolumn{1}{c|}{} &
\multicolumn{1}{c|}{\multirow{2}{*}{\cpar{1.1cm}{from program}}} &
\multicolumn{1}{c|}{\multirow{2}{*}{\cpar{1.1cm}{internal}}} &
\multicolumn{1}{c|}{\multirow{2}{*}{\cpar{1.1cm}{from program}}} & \multicolumn{2}{c|}{internal}  &
\multicolumn{1}{c|}{\multirow{2}{*}{\cpar{1.1cm}{distinct}}} &
\multicolumn{1}{c|}{\multirow{2}{*}{\cpar{1.45cm}{write to\\output args}}} &
\multicolumn{1}{c|}{\multirow{2}{*}{\cpar{1.1cm}{avg \#\\ of args}}}  
 & \multicolumn{2}{c|}{program code}  & \multicolumn{1}{c|}{internal}  &  \multicolumn{2}{c|}{syscalls (int.)}\\
\cline{1-1} \cline{5-6} \cline{10-14}
{\bf Subject} &  & &  & \multicolumn{1}{c|}{tail call} & \multicolumn{1}{c|}{normal} &  &  & & {onEntry} & {onExit} & {onEntry} & {enter} & {exit} \\ 
\hline
APT28 & 0 & 408\,045 & 50\,577 & 1\,934 & 1\,153\,200 & 130 & 29 & 3.20 & 14.38 & 15.76 & 3.16 & 3.28 & 2.33 \\
BlackSquid & 0 & 12\,172 & 4\,667 & 988 & 55\,715 & 151 & 38 & 2.82 & 17.42 & 17.81 & 14.27 & 10.76 & 3.71 \\
Furtim & 88 & 1511 & 541 & 371 & 2\,365\,887 & 71 & 25 & 2.49 & 16.53 & 30.79 & 2.69 & 3.61 & 4.27 \\
Gootkit & 0 & 3\,068 & 4\,737 & 4\,478 & 31\,507 & 79 & 23 & 1.37 & 5.61 & 8.27 & 8.69 & 4.17 & 2.86 \\
Gozi-ISFB & 19 & 1\,509 & 13\,449 & 11\,019 & 22\,180 & 75 & 28 & 1.54 & 5.60 & 6.45 & 3.18 & 4.13 & 9.45 \\
Grobios & 4 & 419 & 225 & 144 & 1\,275 & 27 & 10 & 2.97 & 19.18 & 27.17 & 5.39 & 6.36 & 2.40 \\
Olympic & 0 & 1\,129 & 434 & 298 & 4\,726 & 64 & 26 & 3.26 & 18.38 & 23.36 & 4.59 & 8.22 & 16.83 \\ 
SmokeLoader & 15 & 485 & 49 & 27 & 1\,019 & 28 & 10 & 3.94 & 29.50 & 21.20 & 9.86 & 5.39 & 2.53 \\
Softpulse & 1 & 1\,552 & 1\,163 & 628 & 10\,702 & 83 & 26 & 2.82 & 15.37 & 20.65 & 8.61 & 6.39 & 2.44 \\
Swisyin & 0 & 7\,058 & 1\,392 & 451 & 81\,456 & 22 & 8 & 4.38 & 27.52 & 20.76 & 3.47 & 17.17 & 1.83 \\
Untukmu & 0 & 105\,646 & 23\,978 & 21\,195 & 4\,459\,691 & 25 & 8 & 2.19 & 10.63 & 11.29 & 2.51 & 3.46 & 1.92 \\
\hline
7zip & 0 & 28\,398 & 5\,922 & 294 & 139\,152 & 112 & 26 & 3.80 & 24.36 & 23.37 & 4.55 & 3.89 & 2.41 \\
BitTorrent & 0 & 113\,742 & 268\,804 & 109\,608 & 913\,214 & 366 & 109 & 2.73 & 16.44 & 16.87 & 4.98 & 6.83 & 3.26 \\
Chrome & 1\,821 & 263\,839 & 1\,586\,718 & 684\,236 & 755\,054 & 398 & 145 & 2.99 & 19.68 & 21.56 & 8.83 & 9.21 & 2.64 \\
Foxit Reader & 2 & 150\,490 & 946\,903 & 205\,319 & 818\,568 & 396 & 93 & 3.45 & 17.78 & 19.69 & 4.33 & 5.23 & 2.12 \\
Notepad++ & 0 & 315\,440 & 2\,955\,873 & 1\,645\,034 & 725\,638 & 231 & 36 & 2.04 & 8.63 & 10.57 & 3.90 & 3.15 & 2.24 \\
TeamViewer & 0 & 307\,126 & 489\,341 & 52\,778 & 1\,795\,308 & 328 & 87 & 3.36 & 21.75 & 23.95 & 4.22 & 6.75 & 7.80 \\
\hline
\end{tabular}
\end{footnotesize}
\end{adjustbox}
\vspace{1.5mm}
\caption{API calls recorded on complex malware samples and common productivity programs.\label{tab:sc}}
\end{table*}
%

\subsection{Comparison with Other API Tracing Systems} 
\label{ss:comparison}
We can now detail the comparison depicted in Table~\ref{tab:tools}.

Commodity monitoring products (first three listed entries) introduce classic artifacts \chal{1} and have other limitations. In terms of recall~\chal{2} no one can catch direct syscalls (\twopie{70}), then API Monitor misses also calls to API solved without using the IAT (\twopie{40}), while SpyStudio hooks only a selected number of APIs used deep down in the software stack (\twopie{40}). The three systems have accurate but limited (\twopie{70}) prototypes \chal{3} (especially SpyStudio \twopie{40}), and can trace output arguments \chal{4} as their stubs wrap API returns (\twopie{100}). Discarding internal calls \chal{5} is left to the user  (\twopie{0}). Processes are traced only as a whole, and users have to add child and injected processes \chal{6} to the monitoring manually (\twopie{0}) or using filters in the WinAPIOverride case (\twopie{70}).

Both drltrace and PyREBox fare well with artifacts \chal{1} since they use DBT for instrumentation: drltrace builds on DBI with DynamoRIO, while PyREBox uses whole-system QEMU-TCG emulation.

drltrace looks up EATs to hook API prologues \chal{2} like we do (\twopie{100}). As limitations, it supports fewer APIs than other systems \chal{3} (\twopie{40}), does not trace return values and output arguments \chal{4} (\twopie{0}), and can only follow child processes \chal{6} (\twopie{40}). It has automatic filtering capabilities for internal calls \chal{5} by whitelisting the text and heap regions (we mentioned the pitfalls of this approach in \textsection\ref{ss:relevant-calls-units}), but does not expunge tail transfers used for internal calls (\twopie{70}).

PyREBox interposes on every \verb|call|, \verb|jmp|, and \verb|ret| instruction to hook API entry events \chal{2} (\twopie{100}): however, as discussed in \textsection\ref{sss:api-entry-events} this can add important overhead to the (already high) one from full-system emulation in QEMU~\cite{decafplusplus-RAID19}. PyREBox has a remarkable collection of prototypes \chal{3} (\twopie{100}) that we borrow, and logs output arguments by hooking \verb|ret| instructions in the execution \chal{4} (\twopie{100}). For internal calls users may only encode manual filtering policies for calls across specific pairs of libraries \chal{5} (\twopie{0}). Finally, for derived flows \chal{6} PyREBox follows child processes and conservatively monitors any process that the program interacts with using \verb|NtOpenProcess| (\twopie{70}), while our solution is less noisy as it tracks only injected threads, ignoring the normal activities from legitimate threads of a victim process.


\section{Evaluation}
\label{se:validation}
This section evaluates the capabilities, performance, and security of \scozzertracer{}. Our tracers run on Pin 3.15 and Xen 4.12 (DRAKVUF commit 376c03d), respectively. Users select on startup which DLLs or individual APIs to monitor: we support 446 syscalls and \textasciitilde19K APIs from 194 DLLs.



\subsection{Validation} 
We initially tested our implementations using system utilities and programs shipped with Windows as they use heterogeneous, numerous APIs and occasionally make syscalls. We then collected and analyzed deterministic logs for:
\begin{itemize}
\item the extensive conformance tests~\cite{wine} of the Wine emulator, which cover \textasciitilde10K Windows APIs \chal{3, C4, C5};
\item tools popular in the context of malware sandbox testing like Al-Khaser~\cite{alkhaser} as they use many low-level, seldom undocumented primitives \chal{2, C3};
\item controlled synthetic programs that we shielded with state-of-the-art executable protectors \chal{1, C2}; some exercised derived flows using injection patterns \chal{6}.
\end{itemize}

\noindent
The outcome of testing backed our expectations for \chal{1-6}.





\subsection{Capabilities}
For the evaluation we studied 11 complex malware samples analyzed in~\cite{bluepill-tifs} for their assorted anti-analysis patterns and 6 classic productivity programs. We monitored APIs from 12 popular DLLs that cover different Windows features: {\verb|advapi32|}, {\verb|crypt32|}, {\verb|gdi32|}, {\verb|iphlpapi|}, {\verb|kernel32|}, {\verb|kernelbase|}, {\verb|ole32|}, {\verb|oleaut32|}, {\verb|shell32|}, {\verb|user32|}, {\verb|wininet|}, and {\verb|ws2_32.dll|}. 

To collect the figures reported in Table~\ref{tab:sc} we used an Intel i9-8950HK CPU, 3 GB of RAM, Windows 7 SP1 build 7601 32-bit, and strategy (b) for handling API exits. The DBI and VT-based variants yielded consistent results in the events they recorded; we observed no significant changes when repeating the experiments under Windows 10.

To allow for an indirect quantitative comparison with prior tracing solutions, the table comes with dedicated columns for the respects where those would struggle most (i.e., internal DLL calls and syscalls, output arguments) under the fictitious assumption that they survived the anti-analysis provisions from malware \chal{1, C2, C6}.

The collected figures back our claim on the importance of distinguishing calls originating in program code from internal ones \chal{5}, as the latter may be orders of magnitude more numerous. 
For DLL APIs we qualify internal calls as {\em normal} or {\em tail call} invocations: the ability to discard tail calls in {\em onEntry} (line~\ref{algline:internal}) proved useful, as those can be as numerically relevant as calls from program code (e.g., {\em Gootkit} and {\em Gozi-ISFB}).
Syscalls from program code were few compared to those made within DLL code, yet they can reveal interesting details: for instance, for the {\em Furtim} malware they proved vital for analysts to understand its adversarial strategies according to~\cite{Furtim16,bluepill-tifs}. 

Table~\ref{tab:sc} also reports how many distinct DLL APIs program code invoked, how many of them had output arguments, and the average number of arguments of all kinds from their prototypes \chal{3, C4}. Output arguments turned out relevant in practice, as they were present in 15--40\% of the APIs that we observed.

\subsection{Analysis Time}
We also measured the time spent in executing the monitoring callbacks. The numbers shown in Table~\ref{tab:sc} refer to analysis code only, as probe insertion is a well-studied problem in DBI and VT-based research~\cite{DynamoRIO-CGO03,SPIDER-ACSAC13,Estrada-EDCC15} that leaves us little optimization room. The average processing time for DLL APIs called from program code was 5-31 \textmu{}s for each API entry or exit event, with a high correlation with the number of arguments to process (0.94 Pearson correlation between the {\em onEntry} processing time and the average number of arguments handled by \textit{parseArgsOnEntry}), and with a minor role being played by variable-size arguments.

Filtering out internal calls is cheaper than a logging: {\em onEntry} took 3--15 \textmu{}s to terminate after the range (line~\ref{algline:restrict}) or the tail call (line~\ref{algline:internal}) checks; {\em onExit} executed faster ($<$1 \textmu{}s) likely due to locality effects, and we omit its figures since hardly significant. For syscalls we report figures only for internal ones as they are dominant; note that enter analysis includes the verification step for the return address (\textsection\ref{ss:syscalls}). The additional cost for logging arguments in case of syscalls from program code was around a couple dozens \textmu{}s.

\subsection{Security}
We conclude our evaluation by discussing security aspects of the design and its implementations.

As threat model for the transparency challenge \chal{1}, in \textsection\ref{sss:threat-model} we considered introspective attempts on the executable image in memory (T1) or the contents of DLLs (T2): both of our variants successfully pass those tests. The analysis code and data structures are kept separate by design from the monitored application: in the VT-based scenario they reside in the VM monitor, and for Pin we use the shielding techniques of~\cite{SoK-DBI}. We make no visible changes to the monitored program or DLLs.

Let us consider a sophisticated malware scenario where an adversary aware of the inner workings of SNIPER tries to either \textit{evade} or \textit{break} its tracing process. By evasion we mean that the program makes a call that SNIPER misses or erroneously discards as internal. Note that our instrumentation is exhaustive in terms of recall and copes with derived execution flows. Thus to trick the tracer the only option we foresee for an adversary is a TOCTTOU (time of check to time of use) attack: pushing at API call time a fake return address  on the stack falling in the return ranges blacklist, and from a concurrent thread replacing it with the intended return address before the API terminates. While the reliability of such an attack is yet to be explored in practice, strategy (a) for API exits would come to the rescue: the intended return address has to be visible when an API reaches one of its exit point, so we could use that moment to decide whether or not to log the call.

By breaking the tracing process we mean instead that the adversary tries to make SNIPER crash or to subvert its control flow. As analysis code and data are in regions distinct from program ones, the adversary should aim for data-handling bugs by feeding poisonous data to the tracer. We picture two avenues: exhausting the shadow stack or targeting print helpers for argument types. We rule out the first possibility because, irrespectively of the size chosen for the stack (we also remark that our implementations use a resizable C++ vector), constructing an arbitrarily long sequence of nested API calls is not plausible, plus we enforce a policy that discards internal calls. As for arguments, we only copy them: primitive types have fixed size and we make conservative provisions for variable-length data (\textsection\ref{sss:parameters}). We thus foresee no direct way to break the print helpers either.




\section{Discussion} 
\label{se:discussion}
The design points we recommended do not make use of primitives that are exclusive to specific instrumentation technologies. For this reason they are amenable to different instrumentation solutions: for \scozzertracer{} we chose\footnotemark{} two technologies that can cope well with the threat model of \chal{1} and the other requirements \chal{2-6} of \textsection\ref{ss:challenges}.

The design does not address evasions targeting the peculiarities of the underlying instrumentation technique. For this well-studied problem the implementation can resort to existing mitigations, such as patching the Time Stamp Counter in the VM monitor upon VM exit events~\cite{Brengel-DIMVA16}, or hiding artifacts of a DBI runtime as we did by using the mitigation library of~\cite{SoK-DBI}.

Kawakoya et al. in~\cite{APIChaser13} use taint analysis for an adversarial model for API monitoring where an attacker can evade hooks by emulating with own instructions the initial portion of an API before jumping in the middle of its canonical implementation. We may cope with popular forms of such {\em stolen code} attacks by moving the hook from the initial instruction of an API to a later basic block (for instance, one that post-dominates the entry block in the control flow graph) where arguments are still visible. The authors also consider code injection attacks to elude monitoring using other processes: we tackled this surface by tracking child processes and remote threads. The implementations can be extended to recognize new exotic injections~\cite{Injection-BHUSA19} as \scozzertracer{} would capture the API calls needed to mount them.

Our system is deceived by attacks against API name resolution based on non-standard loading of DLLs. One described in~\cite{APIChaser13}---and countered by the authors using disk-level taint analysis---consists in copying a system DLL to a non-standard path and altering its exported symbols before loading it. A more fragile and conspicuous variant is copying from a loaded DLL the code of individual APIs to program RWX memory; note that internal calls may still leak, as the outermost ones would be seen as from program code. In a more complex attack~\cite{StealthLoader17}, the attacker reimplements the Windows loader and recursively rewires every import referencing other DLLs to use stealth copies of such libraries, so that calls to ``standard'' API functions are never made in the program. As future work we are thinking of exploring the countermeasures suggested in~\cite{StealthLoader17} to extend our system to cope with these orthogonal attacks, which however are also challenged by Windows mitigations like Arbitrary Code Guard and  Code Integrity Guard~\cite{msdn-acg-cig}.

\footnotetext[6]{Instrumenting QEMU for whole-system analysis did not seem appealing: VT-based schemes execute at native speed and bring fewer artifacts. We foresee however no major obstacle to implement our design using it: we could borrow the infrastructure to hook specific instruction addresses from popular QEMU-based projects like DECAF++~\cite{decafplusplus-RAID19}, while our VMI component for tracking threads and processes would only need small adaptations to QEMU.}


\section{Other Related Works}
For dynamic analysis of binary programs researchers have used for a long time designs operating alongside the object of the analysis. In the context of monitoring the interactions with environment, systems of this kind have ranged from operation-specific tracers (e.g., \cite{regmon}) to full-fledged sandboxes. A common strategy was to patch the functions of interest~\cite{Egele-CSUR08}. Microsoft Detours~\cite{Detours-USENIX99} offered general API hooking primitives based on trampolines to invoke a user-defined function, for instance to sanitize sensitive arguments. Similarly, in its day the pioneering CWSandbox~\cite{cwsandbox} replaced the first 5 bytes of each API under observation with a trampoline to an analysis callback. The nowadays popular Cuckoo Sandbox uses similar conspicuous trampolines to monitor about 320 API functions for its analyses~\cite{cuckoo-hooks}.




Code changes and artifacts introduced by such approaches have worried researchers and practitioners already for some time~\cite{Egele-CSUR08}. Garfinkel and Rosenblum in a seminal work~\cite{VMI-NDSS03} proposed to move an intrusion detection system from the guest to the VM monitor, using VMI techniques to inspect the guest with better transparency and isolation. VMI was later adopted in many other scenarios, first and foremost malware analysis and memory forensics.

Ether~\cite{Ether-CCS08} pioneered low-artifact malware analysis with a system based on VT extensions with syscall tracing capabilities: to intercept syscalls Ether used a bogus value in both the \verb|SYSENTER_EIP_MSR| register and the interrupt descriptor table entry 0x2E so to cause a page fault whenever the guest triggered a syscall using \verb|sysenter| or \verb|int 2e|. SPIDER~\cite{SPIDER-ACSAC13} then brought new capabilities to the table with invisible breakpoints for instrumenting selected instruction addresses (\textsection\ref{ss:instrumentations} and \textsection\ref{sss:hook-api-entry}), presenting two case studies (attack provenance and confidential data acquisition) based on manual hook placement.

To the best of our knowledge, our article is the first work to identify and propose solutions for the challenges that arise when trying to monitor the vast universe of APIs offered by Windows. It also represents the first attempt to extend VT-based monitoring to arbitrary user-space API calls in an automated, general-purpose manner.

Our system shares similarities with hprobes~\cite{Estrada-EDCC15}, a framework that uses VT extensions for warm hook insertion in user space. The work discusses three software dependability case studies: an emergency exploit detector, a watchdog, and an infinite-loop detector. To insert hooks hprobes overwrites instructions of interest with {\tt int3} so as to have the VM monitor kick in and carry out the analysis. We find hprobes to serve a purpose orthogonal to ours, as it means to back generic, user-supplied analyses for specific events. Unlike the invisible breakpoint from SPIDER, hprobes makes no provisions for hiding code changes, so its technique is not transparent to checksumming attempts \chal{1}.

This limitation is shared by designs for secure hook insertion inside a VM with OS modifications (e.g.,~\cite{Lares-SP08, SIM-CCS09}), which are alternative to invisible breakpoints. Recent developments in this area~\cite{SeCage-CCS15, ShadowMonitor-RAID18} feature more efficient isolation using the {\tt VMFUNC} feature of VT extensions but introduce distinguishable code artifacts. Lately, the OASIS~\cite{OASIS-SP21} system has made promising improvements in this direction.


We conclude by discussing DBT systems. Those are a popular choice for security analyses that require fine-grained instrumentation capabilities, such as tracking instructions by type (e.g., if they read memory) instead of by address, or performing substantial code modifications. DBT systems usually offer better transparency and flexibility than binary patching~\cite{SoK-DBI}, although they may incur emulation artifacts~\cite{MPR-TOSEM13}.

A recent work~\cite{DeGoer-ACSAC18} uses DBI for real-time function call detection involving own functions of a program. Unlike APIs and \verb|.edata|, entry points for those functions are not declared in the executable. The authors show how to scrutinize control transfer instructions to identify function calls reliably. We believe it would be interesting to compare this approach in terms of recall and efficiency with a solution combining our design for on-entry hooks with recent advances for function detection in binaries~\cite{andriesse-eurosp-2017}.

\section{Conclusion}
API monitoring is a valuable technique in many research scenarios. In this article we identified and proposed solutions for key challenges towards robust API monitoring, a task for which both commercial and research systems reveal several shortcomings. We discussed how to build tracing solutions for Windows binaries and their multifaceted universe of challenges, suggesting general design points amenable to different instrumentation technologies. Our techniques are general: they make no assumption on how a program is compiled or obfuscated, but only on the calling conventions in use. Thus they may be applied also to other systems such as Linux and MacOS with adaptations for the structure of libraries and their loading. We also detailed the first general-purpose tracer that builds on VT extensions, popular today for their efficiency and transparency. 

\bibliographystyle{IEEEtran}
\bibliography{bibliography}

\clearpage

\appendices
\section{Additional Material}
\label{apx:a}

In \textsection\ref{ss:design-precision} we mentioned that when chasing return addresses with strategy (b) to hook API exit events, the instruction corresponding to the return address for some API call may be a join point in the control flow graph of the caller.

If we insert a hook there and do not remove it after the call terminates (for instance, because hook deletion brings overhead that we wish to avoid), the analysis callback must distinguish whether it is intercepting a real API exit event.

In the example below, taken from the 32-bit {\tt calc.exe} shipped with Windows 7 SP1 64-bit (file version 6.1.7601.17514), we instrumented the instruction located at address {\tt 10020cf} when we first intercepted the call to the {\tt LocalFree} API (\verb|kernel32.dll|) from its enclosing function. However, subsequent invocations of the latter eventually reach this address also when coming from another basic block, namely the entry block, which does not end with an API call. The logic of the analysis has to discard these events: in fact, our implementation will find no valid shadow stack entry for it. We found other instances of this pattern in {\tt calc.exe} (e.g., at addresses \verb|100367e|, \verb|100aaba|, and \verb|100cec3|) and several other Windows utilities.

\begin{figure}[hb!]
\begin{center}
\includegraphics[width=0.6\linewidth]{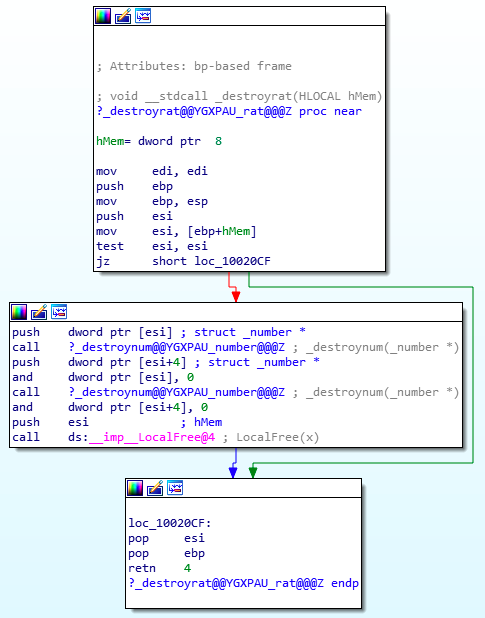}
\end{center}
\vspace{-6mm}
\caption{The instruction at address {\tt 10020cf} in {\tt calc.exe} is a join point in the control flow graph of its enclosing function: it can be reached either by a conditional jump from the entry basic block of its function or as a fall-through for the call to the {\tt LocalFree} API function.
\label{fig:cfg-join-point}}
\end{figure}

\end{document}